\documentclass{aa}
\usepackage{graphicx}

\def\ssim{\setbox0=\hbox{$\propto$}%
\setbox1=\hbox{$<$}\dimen0=\ht1%
\advance\dimen0by-1.2pt\,\lower.6\dimen0%
\copy0\kern-\wd0\raise.4\dimen0\copy1 \,}

\def\gsim{\setbox0=\hbox{$\propto$}%
\setbox1=\hbox{$>$}\dimen0=\ht1%
\advance\dimen0by-1.2pt\,\lower.6\dimen0%
\copy0\kern-\wd0\raise.4\dimen0\copy1\,}

\def\lambdab{\lambda\mkern-9mu\lower1.2pt\hbox{$\mathchar'26$}}%

\begin{document}
 \title{Stellar rotation: Evidence for a large horizontal turbulence
 and its effects on evolution}

\author{Andr\'e Maeder }

     \institute{Geneva Observatory CH--1290 Sauverny, Switzerland\\
              email:  Andre.Maeder@obs.unige.ch
                  }

   \date{Received / Accepted }

\abstract{ We derive a new expression for the coefficient $D_{\mathrm{h}}$ of diffusion by
horizontal turbulence in rotating stars. This new estimate can be up to  two orders of
magnitude larger than given by a previous expression. 
As a consequence the differential rotation
on an equipotential is found to be very small, which 
reinforces Zahn's hypothesis of shellular rotation.  
The role of the so--called $\mu$--currents, as well as the driving of circulation,
are reduced by the large horizontal turbulence.
Stellar evolutionary models for a 20 M${\odot}$ star are calculated with the
new  coefficient. The new and large $D_{\mathrm{h}}$ tends to limit the size of the convective core
and at the same time it largely favours the diffusion of helium and nitrogen to
the surface of rotating OB stars, a feature supported by  recent observations. 
  }

\titlerunning{Horizontal turbulence in stars}   

   \maketitle

%

\section{Introduction}

Differential rotation usually creates  turbulent motions 
 due to shear in rotating 
stars. In a normally stable radiative layer, the turbulence is
usually  much stronger
in the horizontal direction 
  perpendicular to gravity than in the vertical
direction (Zahn \cite{Zahn92}). The reason is that in the vertical direction the stable
 temperature gradient  needs stronger forces to be overcome
than in the horizontal direction where, in principle, no forces
are opposed to the motions. An  argument in favour of these 
intense
horizontal meteorological--like turbulent motions is given by the
study  of the solar tachocline by Spiegel \& Zahn (\cite{spiegzahn92}).
 The tachocline is the transition zone between the rigid rotation in the radiative
interior and the external convective zone, where rotation varies
with latitude. Spiegel \& Zahn show that if the horizontal turbulence 
is intense, then the tachocline is very thin as supported by 
helioseismological observations.

The global result of the transport of the fluid elements by  the horizontal turbulence
is represented by a coefficient of viscosity $\nu_{\mathrm{h}}$. This coefficient 
is a very
important  parameter in the physics of rotating stars in several respects:

--1. If strong enough, the horizontal 
coupling expressed by the coefficient  $\nu_{\mathrm{h}}$ makes
the angular velocity $\Omega$ nearly constant on isobaric surfaces
(cf. Zahn \cite{Zahn92}). In this case, the angular velocity
$\Omega$ is constant on shells and the rotation law is said 
``shellular'' by Zahn. If this is the case, the equations of stellar
structure are greatly simplified, because they  depend on one coordinate only
(which is not just the lagrangian coordinate M$_{\mathrm{r}}$,
 but which has to be defined
in an appropriate way). This enables
us to keep a 1--D equations  scheme for stellar structure (cf. Kippenhahn \& Thomas \cite{kipp70};
Endal \& Sofia \cite{endsof76}). In the case of differential shellular rotation,
Meynet \& Maeder (\cite{MMI}) have shown that the scheme usually employed is
incorrect, but that a consistent 1--D scheme may still be defined.

--2. The various mixing processes of chemical elements play
a major role in massive star evolution (cf. Heger et al. \cite{heger1};
Heger \& Langer \cite{heger2}; Meynet \& Maeder \cite{MMI}).
The horizontal turbulence reduces very much
the efficiency  of  vertical transport of elements by meridional circulation (Chaboyer
\& Zahn \cite{chabzahn92}). This  enables us  to understand in a consistent way
why the vertical transport of  chemical elements  by the circulation is
 much smaller than the vertical transport of  angular momentum by
 circulation.  This is a clear constraint which results from solar observations
 (Chaboyer  et al. \cite{chab95a,chab95b}),
 as well as from the observations of massive stars 
 (Maeder \& Meynet \cite{MMI}).

--3. The horizontal turbulence  was generally  ignored in the
treatment of  meridional circulation or of  shear mixing.
However,  recent developments  (Chaboyer \& Zahn \cite{chabzahn92};  Maeder \& Zahn \cite{MZIII}; 
Maeder \& Meynet \cite{MMVII}; Br\"{u}ggen \& Hillebrandt \cite{bruggen})  show that the horizontal diffusion
by turbulence may 
also intervene in   the expressions of the transport of chemical
elements by  meridional circulation, of the 
 circulation velocity, of the diffusion coefficient by
 shear mixing, of  the heat transport, etc... Interestingly enough, 
 the numerical convergence
 of the 4th order scheme of differential equations  expressing
 the transport of angular momentum  and meridional circulation appears
  to be sensitive  to the value of diffusion coefficient of
 horizontal turbulence.

We do not consider here the effect of the magnetic field
(cf. Spruit \cite{spruit}), which may
also play a role in the transport of angular momentum.
In Sect.2, we examine the reasons which demand  a new estimate
of $\nu_{\mathrm{h}}$. In Sect. 3, we derive a new expression for 
$\nu_{\mathrm{h}}$ and some numerical estimates.
Sect. 4 provides a discussion of the results.

\section{Reasons for a new estimate of the horizontal turbulence}

The usual expression for the coefficient $\nu_{\mathrm{h}}$ of 
viscosity  due to horizontal turbulence and for the coeffcient $D_{\mathrm{h}}$ of
horizontal diffusion, which is of the same order, is, 
according to Zahn (\cite{Zahn92}; Eq.(2.29)),

\begin{equation}
D_{\mathrm{h}} \simeq \nu_{\mathrm{h}} =
\frac{1}{c_{\mathrm{h}}} r \;|2V(r) - \alpha U(r)| \,\, ,
\end{equation}

\noindent
where $r$ is the appropriately defined eulerian coordinate 
of the isobar (Meynet \& Maeder \cite{MMI}). Apart from the case
of extreme rotational velocities, the parameter $r$ is close to  the average 
radius of an isobar, which is the radius at $P_2 (\cos \vartheta) =0$,
namely for $\vartheta = 54.7$ degrees. 
$U(r)$ is the vertical
component of the velocity of meridional circulation,
$V(r)$ the horizontal component, 
$\alpha = \frac{1}{2} \frac{d \ln r^{2} \overline{\Omega}}
{d \ln r}$ and
$c_{\mathrm{h}}$ is a constant of  order of unity or smaller. 
This equation was derived assuming  that the 
differential rotation (as defined  by the ratio $\frac{\Omega_2(r)}
{\overline{\Omega}(r)}$ in  Eq.(4) below) on an isobaric surface  be small compared 
to unity. Indeed, there are 
several difficulties suggesting us to reconsider 
the expression for $\nu_{\mathrm{h}}$:

-- The first reason  why  the above expression is not satisfactory
has been given by Zahn (\cite{Zahn92}) and it is related to the
way Eq.(1) has been obtained. If we write the differential
rotation at a colatitude $\vartheta$ as
\begin{eqnarray}
\Omega(r,\vartheta) = \overline{\Omega}(r)+	
\widehat{\Omega}(r,\vartheta) \quad \mathrm{with}\\ [2mm]
\widehat{\Omega}(r,\vartheta)=\Omega_{2}(r) \, P_{2}(\cos \vartheta) \, ,
\end{eqnarray}

\noindent
where $P_{2}(\cos \vartheta)$ is the Legendre polynomial of second order,
 we find that the  differential rotation  is a constant
(cf. Sect. 2.6 in Zahn \cite{Zahn92}), 
\begin{equation}
\frac{\Omega_{2}(r)}{\overline{\Omega}(r)} = \frac{c_{h}}{5}  \, .
\end{equation}
\noindent
 This ratio is  obtained when we use a coefficient $\nu_{\mathrm{h}}$
given by Eq.(1) together with the expressions for the horizontal transport
of angular momentum.
This ratio is smaller than unity, but,
as noted by Zahn (\cite{Zahn92}), there is no reason for 
 the amount of differential rotation being constant with $r$.
On the contrary, the importance of differential
rotation should depend on the value of $\nu_{\mathrm{h}}$, 
because the stronger the horizontal turbulence, the stronger is
the homogeneisation of the angular velocity on an equipotential
surface. This suggests that 
 $\frac{\Omega_{2}(r)}{\overline{\Omega}(r)}$ should decrease
 for larger   $\nu_{\mathrm{h}}$.
 Likely, we could also expect that the importance of differential
 rotation varies with the rotation velocity, since the 
 horizontal turbulence is itself generated by rotation.

-- Another point is related to  the numerical models (Meynet \& Maeder, 
 \cite{MMV}; Maeder \& Meynet, \cite{MMVII}).  During the
 course of the evolution, some models  indicate that the coefficient  $D_{\mathrm{h}}$ of
horizontal turbulence,
as given by Eq.(1), is not so much larger than the coefficient of vertical diffusion by shear,
especially at low metallicity $Z$ where $U(r)$ is small
 (see Fig. 6 in Meynet \& Maeder \cite{MMV} and Fig. 2 in Maeder \& Meynet 
 \cite{MMVII}). This is  not very satisfactory for
 the validity of the assumption of shellular rotation. We may remark
 in this context that this assumption would be much better if the diffusion coefficient 
of  horizontal turbulence would be larger.

--  We may also note that
physically the horizontal turbulence results from the
differential rotation, while the meridional circulation with components
$U(r)$ and $V(r)$ results from the disruption of the thermal equilibrium
on an equipotential. These two phenomena are generally, but not necessarily, related.
 An example is the case of a uniformly rotating star.
 There we have no differential rotation, but a breakdown of thermal equilibrium
 occurs unavoidably.

\section{The dissipation and feeding of turbulent energy}

Let us  firstly examine the rate of dissipation of the turbulent energy.
As shown by Zahn (\cite{Zahn92}), we may write the rate of viscous dissipation 
of the energy present in the differential  zonal motions on an isobar as
\begin{equation} 
\delta \dot{\epsilon}_{\mathrm{t}}(r, \vartheta) =\nu_{\mathrm{h}}
\left (\sin\vartheta \, \frac{\partial \widehat{\Omega}}{\partial \vartheta}
 \; \delta \vartheta \right)^{2} \; ,
 \label{rate}
\end{equation}

\noindent
 per mass and time and for an interval of latitude $\delta \vartheta$.
Taking into account Eq.(3), we obtain

\begin{equation} 
\delta \dot{\epsilon}_{\mathrm{t}}(r, \vartheta) =\nu_{\mathrm{h}}
\sin^{2}\vartheta \; \Omega_{2}^{2}(r) \left(\frac{d P_{2} (\cos \vartheta)}
{d \vartheta} \right)^{2} \delta \vartheta^2  \; ,
\end{equation}

\begin{equation} 
\delta \dot{\epsilon}_{\mathrm{t}}(r, \vartheta) = 9 \; \nu_{\mathrm{h}}
 \; \Omega_{2}^{2}(r)  \sin^{4}\vartheta  \cos^{2}\vartheta \delta \vartheta^2  \; .
 \label{dissip}
 \end{equation}

\noindent
The rate of energy dissipation is proportional to 
 the square of the amplitude $\Omega_{2}(r)$ on an equipotential. It is zero at the
 pole and equator and maximum at $P_2(\cos \vartheta)= 0 $.
 
There is an excess of energy   on an isobar due to
 the differential rotation described by Eq.(2) compared to an average rotation. 
The velocity of rotation $v(r,\vartheta)$ on the equipotential
of  average distance $r$ is given by 

\begin{equation}
v(r, \vartheta) = r \, \sin \vartheta  \overline{\Omega}
 + r \; \sin \vartheta \Omega_{2}(r) \; P_{2}(\cos \vartheta) \; .
\end{equation}

\noindent
For an interval of latitude $\delta \vartheta$, the difference of rotational
velocity $\delta v (r,\vartheta)$ due to the latitudinal differential rotation
on the equipotential is 

\begin{eqnarray}
	\delta v (r,\vartheta) =  r \, \sin \vartheta \; \Omega_{2}(r) \frac{d P_{2}(\cos \vartheta)}
	{d  \vartheta}  \delta \vartheta  \nonumber  \\[2mm] 
	= - 3 r \; \sin^{2} \vartheta \cos \vartheta \; \Omega_{2}(r) \delta \vartheta .
\end{eqnarray}

\noindent 
 We express here only in the velocity difference due to the
shear on the equipotential.
The excess of energy  $\delta E_{\mathrm{diff}}(r, \vartheta)$
over an interval $\delta
\vartheta$ due to the differential rotation  in latitude is 

\begin{eqnarray}
\delta E_{\mathrm{diff}}(r, \vartheta) = 
\frac{1}{4} \, \delta v^{2}(r, \vartheta)  \nonumber \\[2mm]
=\frac{9}{4}\, r^{2} \sin^{4} \vartheta \cos^{2} \vartheta \; \Omega^{2}_{2}(r)
\; \delta \vartheta^{2} .
\label{excess}
\end{eqnarray}

\noindent
Now, this small excess of energy over an interval $\delta \vartheta$
will be smeared out in a
 dynamical timescale $\delta t_{\mathrm{diff}}$.
 
 Let us estimate this characteristic  timescale. On the isobar, 
 the differential rotation due to $\Omega_2$ produces 
  a shift
 $\delta \varphi$ in longitude for two fluid elements located 
 at a  difference $\delta \vartheta$ in colatitude in a time interval $\delta t$
 
 \begin{equation}
\delta \varphi = \delta \widehat{\Omega} \, \delta t = \Omega_2 {d P_2 \over d
\vartheta} \;
\delta \vartheta \, \delta t  \; .
\end{equation}

\noindent
 The meridional circulation has an horizontal velocity component ${V}$, which
 is pointing toward the pole in the external layers where $U(r) < 0$.
 This is  due to the Gratton--\"{O}pik
 term, which is the term $- \frac{\overline{\Omega^2}}
 {2 \pi G \overline{\rho}}$ (\"{O}pik \cite{op}) which appears 
 in the equation for $U(r)$ given for example 
 in Eq.(4.29) by Maeder \& Zahn (\cite{MZIII}). Due to the average density
 $\overline{\rho}$ at the denominator, the
   Gratton--\"{O}pik term is largely negative near the surface,
 thus it acts so as to change the sign of the circulation  $U(r)$,
 making it rising in the equatorial plane and descending along the
 polar axis. In the deeper layers, one has generally  $U(r) > 0$,
 which means that the circulation rises along the polar axis and descends in the 
 equatorial plane. Thus, in this case $V$ is pointing toward the equator. 
 The shift in latitude is given by
\begin{equation}
  r \; \delta \vartheta =  V  {d P_2 \over d \vartheta} \;\delta t  \; ,
\end{equation}

\noindent
and this leads to

\begin{equation}
\delta \varphi = {\Omega_2 V \over r} \left({d P_2 \over d
\vartheta}\right)^2
(\delta t)^2 .
\end{equation}

\noindent
The complex motion in $\vartheta$ and $\varphi$ 
due to the differential rotation $\Omega_2$
on the isobar  will tend to smear out the latitudinal energy differences as
discussed above. As a typical dynamical timescale, we take the time
necessary for  this differential motion in $\varphi$ 
to perform $n$ axial rotations. We may  consider an
average of $\delta \varphi(\vartheta)$ over the star

\begin{equation}
\overline{\delta \varphi} = {\Omega_2 V \over r} \; 
(t_{\mathrm{diff}})^2  \int^{\pi \over 2}_{0}
\left({d P_2 \over d \vartheta}\right)^2 \sin{\vartheta} \; d \vartheta \;
 = 2 n \pi  \; .
\end{equation}

\noindent
Thus, we get for the characteristic timescale

\begin{equation}
t_{\rm diff} = \left(\frac{5 n \pi}{3}  {r \over \Omega_2 V}\right)^{1/2} \; .
\label{tdiff}
\end{equation}

\noindent
The numerical factor is of course rather arbitrary.
The  ratio of the energy excess  (Eq.(\ref{excess})) and of the rate 
of  viscous dissipation (Eq.(\ref{dissip})) is of the order of this
timescale and we write

\begin{equation}
\frac{\delta E_{\mathrm{diff}}(r, \vartheta)}{\delta \dot{\epsilon}_{\mathrm{t}}(r, \vartheta)}
  = \; t_{\rm diff} \; .
\end{equation}

\noindent
Using Eqs.(\ref{dissip}) and (\ref{excess}), we obtain  
 for the coefficient of horizontal turbulence $\nu_{\mathrm{h}}$ 

\begin{equation}
\nu_{\mathrm{h}} = \left( \frac{3}{80 n \pi}\;
r^3 \Omega_2 V \right)^{\frac{1}{2}} \;.
\label{nuhprelim}
\end{equation}

\noindent
We may also estimate  $\nu_{\mathrm{h}}$
by dividing the square of 
a  typical lengthscale (of the order of $r$) by the  diffusion timescale
given by Eq.(\ref{tdiff}). We obtain exactly the same functional
dependence in $(r^3 \Omega_2 V)^{1 \over 2}$, with a numerical
coefficient depending on the chosen lengthscale.
 Studying conservation of the angular momentum by taking into account  
the horizontal variations $\Omega_{2}$
of rotation leads to the following relation (Zahn \cite{Zahn92}; Eq.(2.27)),
that relates $\Omega_2$ and $\nu_{\mathrm{h}}$

\begin{equation}
\nu_{\mathrm{h}} \; \Omega_{2}(r) = \frac{1}{5} \; \overline{\Omega}(r)
\; r \left[ 2 V - \alpha U \right] \; ,
\label{zahno2}
\end{equation}

\noindent 
where $\alpha$  is the same as in Eq.(1). This is the expression
discussed in Sect. 2 , which implies that,
if $\nu_{\mathrm{h}}$ is given by an equation like  Eq.(1),
the ratio $\frac{\Omega_{2}}{\overline{\Omega}(r)}$ is a constant.
Now, we   eliminate
$\Omega_2$ between the two equations (\ref{nuhprelim}) and 
(\ref{zahno2}).  This gives for the coefficient of viscosity due
to the horizontal turbulence

\begin{eqnarray}
\nu_{\mathrm{h}} =  A \; r \; \left(r
\overline{\Omega}(r) \; V \;
 \left[ 2 V - \alpha U \right]\right)^\frac{1}{3}  \nonumber  \\ [2mm]
 \mathrm{with} \quad   A = \left( \frac{3}{400 n \pi} \right)^{\frac{1}{3}} \;.
\label{nuh}
\end{eqnarray}

\noindent
For n=1, 3 or 5 $A \approx 0.134, 0.0927, 0.0782$ respectively. 
This expression  can be written  in
the usual form  $\nu_{\mathrm{h}}= \frac{1}{3} \; l \cdot v$ for a viscosity, 
where the appropriate velocity $v$
is a geometric mean of  3 relevant velocities: 
\begin{itemize}
\item A velocity $(2 V - \alpha U)$ as in Eq.(1) by Zahn (\cite{Zahn92}), 
\item The horizontal component $V$ of the 
meridional circulation. 
\item The average local rotational velocity $ r \overline{\Omega}(r)$. This 
rotational velocity is usually much larger than either $U(r)$ or $V(r)$, typically by
6 to 8 orders of a magnitude in an upper Main Sequence star rotating with the average
velocity. 
\end{itemize}

\section{Numerical results}
\subsection{Comparisons and orders of magnitude}

Let us compare the present value of $\nu_{\mathrm{h}}$ to that given by Zahn
(\cite{Zahn92}).  We get the following ratio from Eq.(1) and (\ref{nuh}),

\begin{equation}
\frac{\nu_{\mathrm{h}}(present)}{\nu_{\mathrm{h}}(Zahn)}= \;
A \, c_{\mathrm{h}}
\left({\frac{r \;  \overline{\Omega} \; V } 
{(2V - \alpha U )^{2}} }\right)^{\frac{1}{3}} \;.
\label{nupresnuz}
\end{equation}

\noindent
The quantities $V$ and $U$ have  the same  order of magnitude. The numerical
models below show that typically $ V \approx  \frac{1}{3} U $  and
$(2V - \alpha U ) \approx V$,
thus we have the following order of magnitude,

\begin{equation}
\frac{\nu_{\mathrm{h}}(present)}{\nu_{\mathrm{h}}(Zahn)} \approx
\; A \, c_{\mathrm{h}}
\left(\frac{r \;\overline{\Omega}}{V}\right)^{\frac{1}{3}} \; .
\end{equation}

\noindent
Thus, we see that the ratio of the two estimates of the diffusion coefficient
is equal to  the power $\frac{1}{3}$  of the ratio  of the local  rotational velocity  to
the horizontal velocity  of meridional circulation at the considered level.
Let us consider a 20 M$_{\odot}$ star with an average rotation velocity of 220 km/s at the surface.
At the middle of the MS phase, the vertical component of the meridional
 circulation lies between $3 \cdot 10^{-4}$ and 
$3 \cdot 10^{-3}$ m/s as shown by the models below (see also Meynet \& Maeder
\cite{MMV}). Thus, we typically have
$\frac{\nu_{\mathrm{h}}(present)}{\nu_{\mathrm{h}}(Zahn)}$ of the order of $10^2$
(cf. Fig. 1). Thus, our  estimate of the diffusion  coefficient of the 
horizontal turbulence  is much larger than  the coefficient proposed by
 Zahn(\cite{Zahn92}; Eq.(2.29)) as given by Eq.(1).

Let us now estimate the degree of differential rotation
corresponding to  this value of  $\nu_{\mathrm{h}}$. From Eq.(\ref{zahno2}),
we have with Eq.(\ref{nuh}),
\begin{eqnarray}
\frac{\Omega_2}{\overline{\Omega}(r)}=
 \frac{1}{5 \; A} \left( \frac{(2V-\alpha U)^2}
{r \; \overline{\Omega} V}\right)^{\frac{1}{3}}  
\label{o2o}          
\end{eqnarray}

\noindent
 There is of course no coefficient $c_{\mathrm{h}}$ in this ratio.
Numerically, this is  1/5 of the inverse of the ratio given
by  Eq.(\ref{nupresnuz}), in which the value of 
$c_{\mathrm{h}}= 1$ would be used. This results from  Eq.(1) and Eq.(18) relating 
$\nu_{\mathrm{h}}$  and $\Omega_2$.
Thus, with the  above estimates, we obtain a ratio of about
$\frac{\Omega_2}{\overline{\Omega}(r)} = 2 \cdot 10^{-3}$.
As the value of  $\nu_{\mathrm{h}}$ obtained in this work
is much larger than the value given by Eq.(1), we
see that  quite logically the degree of differential rotation
on an isobar is  much smaller. The present value of the coefficient
reinforces Zahn's  hypothesis of shellular rotation.
We also  notice that the ratio $\frac{\Omega_2}{\overline{\Omega}(r)}$
is larger for slowly rotating stars. This  is quite a consistent feature,
because $\nu_{\mathrm{h}}$ is growing with the velocity of rotation.

\begin{figure}
   \centering
   \includegraphics[width=8.5cm]{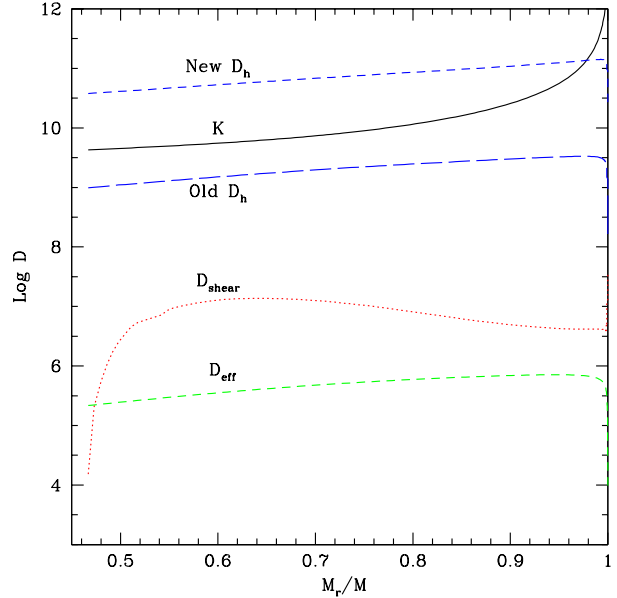}
      \caption{Values of the various diffusion coefficient in the interior of a star model
     of 20 M$_{\odot}$ at beginning  of the MS phase, 
      with an age of $8.577 \cdot 10^4$ yr.
      and central hydrogen content of $X_{\mathrm{c}} = 0.702$.
      That is to say during the initial non--stationary phase of 
      convergence of the rotation, where the velocities $U(r)$ are large. 
      The new   $D_{\mathrm{h}}$ is that given by Eq.(\ref{nuh}) with
      A=0.079. The old $D_{\mathrm{h}}$ is that given by Eq.(1).
       $K$ is the radiative diffusivity. $D_{\mathrm{shear}}$
        is the coefficient of diffusion by shears.  $D_{\mathrm{eff}}$ expresses the diffusion 
        of the chemical elements by meridional circulation with account of the effects 
        of horizontal turbulence.  When not specified, the quantities shown are
        those for the new $D_{\mathrm{h}}$.}
   \end{figure}

\subsection{Test with the evolution of a 20 M$_{\odot}$ model}

In order to examine the consequence of the new coefficient of horizontal diffusion,
we calculate stellar models for a 20 M$_{\odot}$ with composition $X=0.705$
and $Z=0.02$ with the same  physics  as in our recent papers (Maeder \& Meynet
\cite{MMVII}).  The initial rotation velocity is 300 km/s, which corresponds
to average rotation during the MS phase of about 240 km/s.
Several expressions and diffusion coefficients will be
discussed numerically below, let us briefly recall  them.
The vertical component $U(r)$ of the velocity of
meridional circulation velocity is given by

\begin{eqnarray}
{U(r) =  \frac{P}{{\rho} {g} C_{\!P} {T}
\, [\nabla_{\rm ad} - \nabla +
 (\varphi/\delta) \nabla_{\mu}] } }\;  \times  \nonumber  \\[2mm] 
 \left\{  \frac{L}{M_\star} (E_\Omega + E_\mu) \right\} \; .
 \label{U}
\end{eqnarray}

\noindent
 P is the pressure, $C_P$ the specific heat, 
$E_{\Omega}$ and $E_{\mu}$ are  terms depending on the $\Omega$--
and $\mu$--distributions respectively, up to the third order derivatives
and on various thermodynamic quantities (see details in Maeder \& Zahn, 
\cite{MZIII}). The term $E_{\Omega}$ expresses the driving effects of meridional
circulation, while the term $E_{\mu}$ expresses the $\mu$--currents which
tend to inhibit the circulation.
The term $\nabla_{\mu}$ is very important numerically,
 its origin in this expression is more complex than could be thought at first 
 sight. This expression also prevents infinite velocities at the edge 
 of semiconvective zones. The term $E_{\Omega}$ expresses the driving 
 of the circulation by the fluctuations of density due to the
 breakdown of radiative equilibrium. 
 
 The diffusion by shear instabilities is expressed 
 by a coefficient  $D_{\mathrm{shear}}$, namely

\begin{eqnarray}
D_{\mathrm{shear}} =   \frac{ 4(K + D_{\mathrm{h}})}
{\left[\frac{\varphi}{\delta} 
\nabla_{\mu}(1+\frac{K}{D_{\mathrm{h}}})+ (\nabla_{\mathrm{ad}}
-\nabla_{\mathrm{rad}}) \right] }\; \times  \nonumber \\[2mm] 
 \frac{H_{\mathrm{p}}}{g \delta} \; 
\left [ \frac{\alpha}{4} \left(f \Omega{d\ln \Omega \over d\ln r} \right)^2
-(\nabla^{\prime}  -\nabla) \right] \; .
\label{dshear}
\end{eqnarray}

\noindent
where $f$ is a numerical factor equal to 0.8836, $K$ is the thermal diffusivity and
 $(\nabla^{\prime}  -\nabla)$ expresses the difference between the internal
 nonadiabatic gradient and the local gradient (Maeder \cite{MII}).
There is also the coefficient $D_{\mathrm{eff}}$, which expresses 
the contribution of the meridional circulation and horizontal turbulence to the diffusion of 
the elements (Zahn \cite{Zahn92}),

\begin{equation}
D_{\mathrm{eff}} = \frac{|r \; U(r)|^2}{30 D_{\mathrm{h}}} \; ,
\label{deff}
\end{equation}

\noindent
 while the transport of angular momentum by
circulation has to be treated explicitely as an advection.
More details on these various expressions, on the hypotheses leading to
them and on their  domain of validity
can be found in the given references.

 Fig.1 shows the diffusion coefficients  at the very beginning of the MS phase.
There, the situation is non-stationary during 1-2 \% of the MS lifetime,
until the rotation has converged toward an equilibrium profile, (in reality 
a part of this convergence, but probably not the whole,
  may be achieved during the pre-MS phase).
In this temporary stage, $U(r)$ is usually much larger
 (about a few $10^{-2}$ m s$^{-1}$) than later
in the course of evolution, where it is only of the order of a few  $10^{-3}$ m s$^{-1}$ 
 (Meynet \& Maeder \cite{MMV}). 
 We  point out   the much larger value of the new $D_{\mathrm{h}}$
with respect to the old one. 
With the new $D_{\mathrm{h}}$, we see that $D_{\mathrm{eff}}$ is  rather small
with respect to $D_{\mathrm{shear}}$, while with the old $D_{\mathrm{h}}$,
 the coefficient $D_{\mathrm{eff}}$ 
would have been larger than $D_{\mathrm{shear}}$ everywhere,  and in
 particular by several  orders of a magnitude close to the core.  
As to  $D_{\mathrm{shear}}$, the effect is opposite, the new 
$D_{\mathrm{h}}$ makes it bigger since $K$ is replaced by 
$K+D_{\mathrm{h}}$, when the $\mu$--gradient is small.

Fig.2 shows the various diffusion coefficients near
the middle of MS evolution. Interestingly enough, the star shows 3 cells 
of meridional circulation. At the interfaces located at
$M_{\mathrm{r}}\over M$ = 0.535 and 0.950, the nulling of $U(r)$ produces
a kink in the curves of $D_{\mathrm{h}}$, $D_{\mathrm{shear}}$ and $D_{\mathrm{eff}}$.
The outer cell is the Gratton--\"{O}pik cell, due to the lower density in the outer layers.
The main inner cell is the usually dominant cell where $U(r)$ is positive.
There the circulation rises along the polar axis and descends in the equatorial plane,
(thus bringing angular momentum toward the interior). The third
cell close to the core is not a well understood one. It was already present in some curves
of Fig.~4 in Meynet \& Maeder (\cite{MMV}). The velocities here are  very small and
slighty negative. We interpret this third cell as due to a change of the
second derivative of the angular velocity $\Omega$, which influences the
expression of $U(r)$ as given by Maeder \& Zahn (\cite{MZIII}). 
(We also remark a kick in the curve of $D_{\mathrm{shear}}$
at $M_{\mathrm{r}}\over M$ = 0.41; it is produced by variations of
the  nearly vertical gradient of
$\mu$ in some regions.) 
 
\begin{figure}
   \centering
   \includegraphics[width=8.5cm]{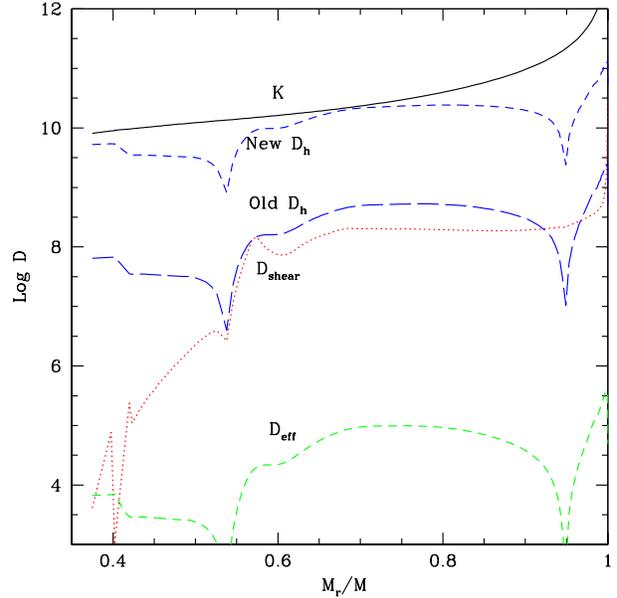}
      \caption{Values of the various diffusion coefficients in the interior of a 
      star model of 20 M$_{\odot}$ at about the middle of the MS phase, 
      with an age of $7.066 \cdot 10^6$ yr. Same remarks about the coefficients
      of diffusion as in Fig.1. }
   \end{figure}

Fig.~2  tells us a lot about the diffusion
coefficients in the stellar interior and their effects:\\
-- As discussed above, 
the new $D_{\mathrm{h}}$ given by Eq.(\ref{nuh}) is larger by about 
2 orders of a magnitude with respect to the old one given by Eq.(1), 
the new value is not far from the thermal diffusivity $K$.\\
-- The new $D_{\mathrm{h}}$  brings some change to $D_{\mathrm{shear}}$.
In regions where the $\mu$--gradient is negligible, the ratio $\frac{D_{\mathrm{shear}}(new)}
{D_{\mathrm{shear}}(old)}$
of the coefficients of shear diffusion  calculated with the new and the old
  $D_{\mathrm{h}}$ behaves like
$\frac{K+ D_{\mathrm{h}}}{K}$. In view of the values in Fig.~2, this means
that $D_{\mathrm{shear}}$ in the outer regions is increased only moderately, currently
 less than a factor of two.  Comparisons with Fig.~6 by Meynet \& Maeder (\cite{MMV})
 confirms the comparable order of magnitude of $D_{\mathrm{shear}}$. \\
-- When $\nabla_{\mu} >> (\nabla_{\mathrm{ad}} - \nabla_{\mathrm{rad}})$,  a situation
which occurs close to the convective core, the ratio   $\frac{D_{\mathrm{shear}}(new)}
{D_{\mathrm{shear}}(old)}$  behaves like   $\frac{D_{\mathrm{h}}(new)}
{D_{\mathrm{h}}(old)}$. This means that 
$D_{\mathrm{shear}}$ is increased by a factor of 100 in the internal
regions close to the core. Such a change should normally strongly favour mixing in the star,
however this is not so much the case, because precisely in the regions close
to the core $D_{\mathrm{shear}}$ is very small due to the very steep 
$\mu$--gradient, which limits the shear diffusion as shown by Eq.(\ref{dshear}). 
Close to the core, $D_{\mathrm{eff}}$ is generally similar or larger than 
$D_{\mathrm{shear}}$ (this was particularly the case when the low 
$D_{\mathrm{h}}$ given by Eq.(1) was used).   \\
-- Contrarily to the case of $D_{\mathrm{shear}}$, $D_{\mathrm{eff}}$ is reduced by an
increase of $D_{\mathrm{h}}$, as is evident from Eq.(\ref{deff}).
 This  can also be seen from a comparison between the present Fig.~2
and   Fig.~6 by Meynet \& Maeder (\cite{MMV}), where much larger
 values of $D_{\mathrm{eff}}$ can be seen.\\
-- Last but not least,  the old  $D_{\mathrm{h}}$ was often
 of the same order as the old $D_{\mathrm{shear}}$ in some  parts
of the star. This was not satisfactory, in view of the hypothesis of shellular
rotation as mentioned in Sect.~2. The new $D_{\mathrm{h}}$, which is much larger
than the new $D_{\mathrm{shear}}$ (cf. Fig.~2),
 solves the problem and makes the hypothesis 
of shellular rotation a much better one as also indicated  by Eq.(\ref{o2o}).
 
   Thus we see  that a larger horizontal turbulence makes $D_{\mathrm{shear}}$ larger 
   and $D_{\mathrm{eff}}$ smaller. The situation is complex, since the 
    relative importance of these two coefficients 
   is not the same throughout the star. $D_{\mathrm{shear}}$ always dominates at some distance
   of the stellar core, while $D_{\mathrm{eff}}$ tends to dominate near the core,
   especially if $D_{\mathrm{h}}$ is small. In addition,
   the ratio of these two coefficients is changing during evolution,
   as seen from Fig.~1  and Fig.~2. Thus, a change of
   $D_{\mathrm{h}}$ affects the evolution of a star in a complex way. 
    These new results now seem kind of very similar to Heger et al.~(\cite{heger1}).
   \emph{In a rough summary,
   we may say that a larger $D_{\mathrm{h}}$ tends to reduce or contain the size of the core, since 
   $D_{\mathrm{eff}}$ which is important close to the core is reduced; at the same time
   the spread of the  processed elements (He and N) up to the surface is larger.}

   \begin{figure}
   \centering
   \includegraphics[width=8.5cm]{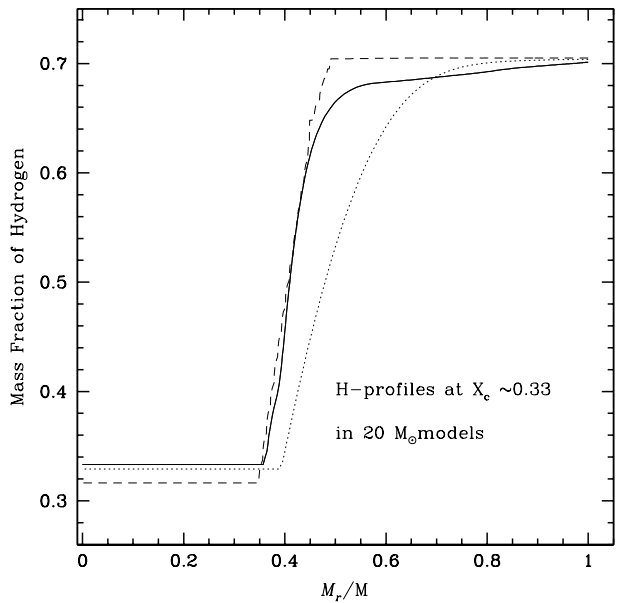}
      \caption{Distribution of hydrogen in models during the MS phase when
      $X_{\mathrm{c}} \approx 0.33$ (cf. Fig.2).  The broken line shows the profile
      for a model without rotation. The continuous line shows the H--profile for a model
      with an initial rotation velocity of 300 km/s
      with new $D_{\mathrm{h}}$, while the dotted line shows the H--profile for a
      rotating model with same rotation velocity  and  the old $D_{\mathrm{h}}$. }
   \end{figure}

A larger horizontal turbulence  $D_{\mathrm{h}}$ also reduces the horizontal
$\mu$--gradients and thus it limits the importance of the so-called
$\mu$--currents introduced by Mestel (\cite{mestel53,mestel65}; see also
Theado \& Vauclair, \cite{thea01} and by Palacios et al, \cite{pal02}). 
 Quantitatively, the term $E_{\mu}$ which expresses the $\mu$--currents contains 
a term $\Lambda=\frac{\tilde{\mu}}{\mu}$, as shown by Eq.(4.30) by Maeder \& Zahn (\cite{MZIII})
and $\Lambda$ itself goes like $(D_{\mathrm{h}})^{-1}$ in a stationary situation
(Eq.(4.40) in above reference). This establishes the relation of $D_{\mathrm{h}}$
with the $\mu$--currents. 

In addition, the horizontal turbulence also contributes
to smear out the temperature and density fluctuations on an equipotential and this
reduces the effects driving meridional circulation. Quantitatively, this is 
expressed by the term containing $D_{\mathrm{h}}$ in the expression of $E_{\Omega}$
in Eq.(4.37) by Maeder \& Zahn (\cite{MZIII}). Thus, globally a higher $D_{\mathrm{h}}$
reduces both the terms driving the meridional circulation as well as the term
which inhibit this circulation. In the numerical example, we see that the values of 
$U(r)$ mentioned above are generally
smaller than those found by Meynet \& Maeder (\cite{MMV}).

Fig.~3 illustrates the  effects  of the  diffusion coefficients $D_{\mathrm{h}}$ 
on the internal distribution of hydrogen. The model with the new and higher values of
$D_{\mathrm{h}}$  has a convective core and a surrounding H--profile which is close 
to that of the non--rotating model, in particular we see that the H--profile close to the core
is much steeper than for the rotating model with the old $D_{\mathrm{h}}$. 
In the outer layers, the H--content of the rotating model with the new 
$D_{\mathrm{h}}$ is lower than for the other two cases
which means than mixing has been more efficient there. These properties
 are quite consistent  with our previous discussion. Indeeed, the higher $D_{\mathrm{h}}$  reduces
the coefficient $D_{\mathrm{eff}}$, which was the largest one close to the core.
This prevents the growth of the core and creates the steep $\mu$-gradient just above it.
Now,  the larger $D_{\mathrm{h}}$ makes $D_{\mathrm{shear}}$  larger outside 
the region of the very steep $\mu$--gradient and this favours mixing in the outer layers.
As a consequence, the enrichments in helium and nitrogen at the stellar surface
are higher. This explains the rather paradoxical result that the model with the
higher $D_{\mathrm{h}}$  has a slightly smaller convective core and at the same
time a larger enrichment in CNO processed elements at the stellar surface.
   
   \begin{figure}
   \centering
   \includegraphics[width=8.5cm]{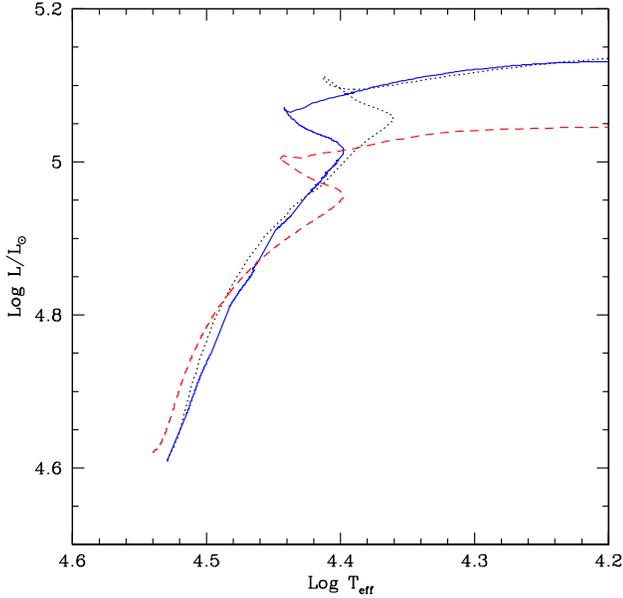}
      \caption{The HR diagram for the MS phase of 3 models of 20 M$_{\odot}$
      at $Z=0.02$. The broken line
      (lower turnoff) is for a non--rotating model. The continuous line
      is for an initial velocity of 300 km/s  with the new  $D_{\mathrm{h}}$ given
       by Eq.(\ref{nuh}). The track  with a dotted line (higher turnoff)
       is for the same initial velocity  with the old  $D_{\mathrm{h}}$ given 
       by Eq.(1).  }
   \end{figure}

Fig.~4 illustrates the tracks in the HR diagram. We see that the model with  the
new (and large)  $D_{\mathrm{h}}$ has a turnoff inbetween that of the model without rotation 
and that of the rotating model with the old coefficient $D_{\mathrm{h}}$ given by Eq.(1). This is
quite in agreement with the H--profiles and the
values of the  mass fractions of the convective cores,
which are 6.8, 7.1 and 7.7 M${\odot}$ when $X_{\mathrm{c}}= 0.33$
 for the model with zero rotation, for the
model with rotation, with  the new 
and  the old $D_{\mathrm{h}}$ respectively. 
As well known, larger cores make MS tracks extend to higher luminosities. We notice
however that the intermediate track with the new $D_{\mathrm{h}}$ is slightly bluer
than an average of the other two tracks would suggest. This is due to the larger enrichments 
of the outer layers in helium, which reduces the opacity and makes the star slightly bluer
and brighter.

\begin{figure}
   \centering
   \includegraphics[width=8.5cm]{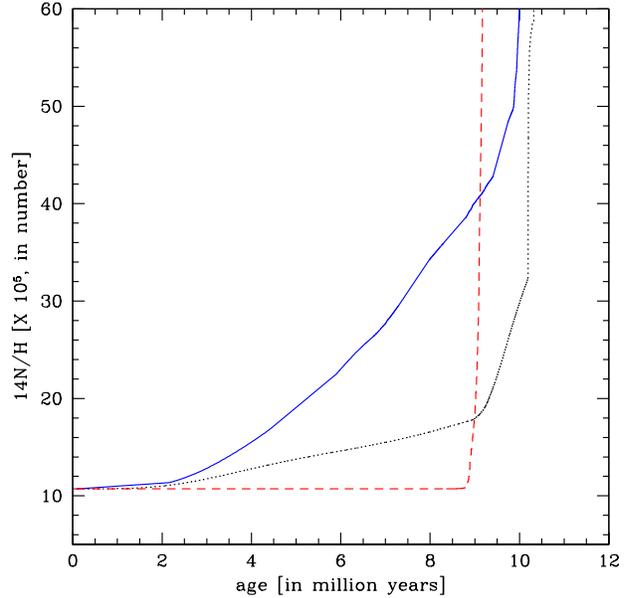}
      \caption{Evolution as a function of time of the abundance ratios 
      $N/H$ at the stellar surface of the 3 models with
      20 M$_{\odot}$ considered.  The meaning of the  lines is 
      the same as in Fig.~3 and 4. The continuous (higher) line
      is that of the rotating model with the new $D_{\mathrm{h}}$.
     }
   \end{figure}

Fig.~5 completes this picture by showing the evolution with time of the ratio 
$N/H$ of the nitrogen to hydrogen ratios at the stellar surface. We see that the
surface enrichment in nitrogen of the model with the new $D_{\mathrm{h}}$
given by Eq.(\ref{nuh}) is larger than the one obtained with old $D_{\mathrm{h}}$
given  by Eq.(1). This is quite consistent with what we have just seen above in
Fig.~3. The larger $D_{\mathrm{h}}$ makes $D_{\mathrm{shear}}$ larger in the outer
layers and the  transport of the new helium and nitrogen to the surface is more
important. This observational consequence is particularly interesting in view 
of the new results by Heap (\cite{heap}), who has shown very high $N/H$ enrichments 
in OB stars up to about 50. Future detailed comparisons
considering carefully the mass, velocities and abundances of OB stars
are necessary to examine whether the  models with the new 
 $D_{\mathrm{shear}}$  are better supported by the observations.

\section{Conclusions}

The following conclusions have been obtained here:

-- By expressing the balance between the energy dissipated by the horizontal 
turbulence and the excess of energy present in the differential rotation on
an equipotential which can be dissipated in a dynamical time, we have
found a new expression for the coefficient of diffusion $D_{\mathrm{h}}$ by 
the horizontal turbulence in rotating stars. This new coefficient is typically larger by a factor 
$10^2$ than the one proposed by Zahn (\cite{Zahn92}).\\
--The differential rotation on an equipotential is found much smaller
so that the hypothesis of shellular rotation by Zahn (\cite{Zahn92})
is reinforced.\\
--A higher horizontal turbulence reduces the importance of the 
$\mu$--currents and also to  a smaller extent the driving of the 
meridional circulation.\\
--Numerical models show in agreement with a physical discussion
that due to the different effects of the horizontal turbulence on the
shears and on the transport of chemicals by circulation, a larger
$D_{\mathrm{h}}$ tends to contain the size of the core and at the same
time to favour the spread of the processed elements up to the stellar surface.\\
--The tracks in the HR diagram obtained with the new and larger $D_{\mathrm{h}}$
for rotating stars are in agreement with the above effects.

It will be  interesting to further explore the consequences of the larger
$D_{\mathrm{h}}$ suggested here for other stellar masses and evolutionary stages.

\begin{acknowledgements}
 I express my thanks to Georges Meynet and Jean--Paul Zahn for their
  useful comments during this work.  Useful remarks by Alexander Heger 
  are also acknowledged with thanks. 
\end{acknowledgements}

\end{document}